# Nanoscale data storage

J. C. Li

*School of Chemistry, The University of Melbourne, Parkville, VIC 3010, Australia*
(Date: Jan/29/2007)

**The object of this article is to review the development of ultrahigh-density, nanoscale data storage, i.e., "nanostorage". As a fundamentally new type of storage system, the recording mechanisms of nanostorage may be completely different to those of the traditional devices. Currently, two types of 'molecules' are being studied for potential application in nanostorage. One is molecular electronic elements including molecular wires, rectifiers, switches, and transistors. The other approach employs nanostructured materials such as nanotubes, nanowires, and nanoparticles. The challenges for nanostorage are not only the materials, ultrahigh data-densities, fabrication-costs, device operating temperatures and large-scale integration, but also the development of the physical principles and models. There are already some breakthroughs obtained, but it is still unclear what kind of nanostorage systems can ultimately replace the current silicon based transistors. A promising candidate may be a molecular-nanostructure hybrid device with sub-5 nm dimensions.**

## 1 Introduction

In previous articles, the issues of ultra-high density data storage are reviewed from the viewpoints of magnetic, optical and electrical recording mechanisms, respectively. Although it has been proposed that silicon based field-effect transistors (FETs) could be scaled down to 10 nm, they will face big challenges with leakage current, power consumption, and random charge/fabrication fluctuations [1]. As predicted by Moore's Law, however, fundamental change is needed in the way present semiconductor-based electronic devices are fabricated by the year 2025, in order to maintain the current rate of increase in the circuit density (doubling every 18 months). In the last century, the transition from one technology to another has occurred several times in information industry. For example, the mechanical relay was replaced by the vacuum tube, which was then substituted by the transistor. Subsequently, the transistor gave way to the current integrated circuit. Therefore, for information storage, we may need to develop nanoscale data storage devices with novel principles, materials, configurations, and fabrication approaches [2].

Today, the world leading semiconductor companies and governments are putting more and more funding into nanotechnology. According to a report of Nanomarkets Co. [3], by 2011 the market for nanoengineered information storage devices will be worth US$65.7 billion. Analysis indicates that much of the present nanotechnology related work on processing and logic is not likely to have a big commercial impact for a few years. Conventional memory and disk systems will continue to exist, and major semiconductor suppliers will continue to hold the dominant position for a few decades. However, new big nanotech companies will be created as the result of innovations in nanoengineered information storage devices and systems. Moreover, with the contribution of nanoengineered storage devices, pervasive, low-cost and high capacity smartcards, sensors, mobile computers and communication devices will become commercially available. This would have a big impact on people's daily life from the perspective of ultra-high efficient network, mobile computing, entertainment and communication devices.

Two approaches have been suggested and applied to fabricate nanoscale devices and systems: top-down miniaturization and bottom-up construction [4,5]. For the top-down case, Feynman outlined in 1959 that: Machines that would make smaller machine, which will be further used to make smaller machines and so on [4]. As an example, researchers have started to build devices at the single-molecule level by using the recently invented scanning tunneling microscope (STM) [6,7,8] and other probe microscopy instruments [9]. In contrast, Drexler suggested a bottom-up path in 1981, in which molecular machines will be constructed via precise position and connection of each molecular building unit in a designated location [5]. For instance, it was argued recently that mono-molecular electronic devices may be possible with integration of wire, switch and/or storage units into a single molecule with memory and data storage functions [10].

With these motivations, scientists from a variety of disciplines are trying to fabricate suitable building blocks for such applications. Prototypes of nanoscale devices have been demonstrated. Briefly, two kinds of 'molecules' are being investigated as active media for nanostorage. One is nanostructured materials such as nanotubes, nanowires, and nanoparticles [11,12,13,14,15,16]. The other is molecular electronic elements such as wires, polymers, dendrimers, and biomolecules [17,18,19,20,21,22]. Fig. 1 shows a schematic flowchart reflecting a possible relationship between these different building blocks and nanostorage device. Hopefully, this research will provide some effective and low-cost ways to realize ultrahigh density data storage with dimensions down to sub-5 nm, i.e., nanostorage. Note that this size falls into the scale of a nanoparticle or a dendrimer molecule.

There is no doubt that we need nanostorage devices in the 'near' future, but it is still not quite clear how to get there [23]. What fabrication approaches and operating principles should be used to realize it effectively and inexpensively? The future nanostorage devices may not be limited to the configuration and the recording media/mechanism of the current devices. So regardless of the recording mechanism, this chapter is mainly focused on the recent development of nanoscale data storage from the viewpoints of device fabrication and characterization. We first discuss the investigation of nanostorage devices on the basis of functional organic molecules. Then we will focus on the memory devices based on dendrimers and biomolecules. Finally, other promising approaches and prospects are briefly discussed.

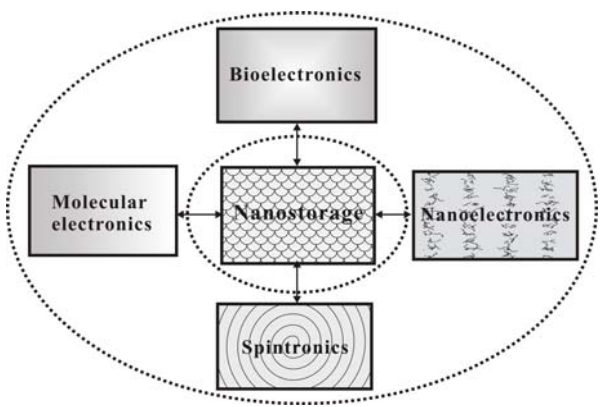

**Figure 1.** Schematic diagram showing a possible correlation between nanostorage and various building blocks.

## 2  Molecular electronics

Molecular electronics is one of the most promising nanostorage approaches, which has the potential to replace current semiconductor based information devices [24,25]. Joseph Weiss discovered in 1942 that charge transfer could happen in certain molecular complexes with electron donor and acceptor molecules [26]. The quantum mechanics of these reactions were elucidated by Mulliken ten years later [27,28]. In 1973, Ferraris and co-workers reported the preparation of a highly conductive organic complex between the electron donor tetrathiofulvalene (TTF) and the electron acceptor tetracyano-p-quinodimethane (TCNQ) with TTF/TCNQ molar ratio of 1:1 (Fig. 2) [29]. When highly purified TTF and TCNQ were co-dissolved in acetonitrile solvent, the 1:1 complex precipitated from the solution. Even when a large excess of TCNQ was used in the synthesis, only the 1:1 complex was obtained. Significantly, the UV spectrum of the complex in hexamethylphosphoramide solvent showed only the presence of the TTF radical and the TCNQ anion, while no absorption peak of neutral TCNQ was observed. The complex crystal has a layered structure with weak interactions between them. Such unique properties make TTF-TCNQ complex an ideal example and impetus for studying highly conductive organic compounds with better structure than that of conducting complexes such as TTF$^+$Cl$^-$ and M$^+$ TCNQ$^-$ [30,31].

Soon after the discovery of the TTF-TCNQ complex, Aviram and Ratner proposed a novel molecular rectifier based on a molecule with structure of donor-bridge-acceptor [32]. As shown in Fig. 2, the molecule was designed based on TTF and TCNQ with a triple methylene bridge, i.e., TTF-σ-TCNQ. The bridge can effectively prevent the interaction between the donor and the acceptor on the time-scale of electronic motion to or from the electrodes in order for the device to function. They examined possible electrical rectification by looking at the current-voltage (I-V) characteristics of a single molecule connected with two metal electrodes. Both the sigma bridge and the two metal/molecule interfaces were treated as thin tunneling barriers. The shift of the device energy levels was then analyzed under forward and reverse bias voltages, respectively. The results suggested that it is possible to design molecules that have a larger threshold voltage for conduction in one direction than for the other direction at an appropriate device structure and bias voltage.

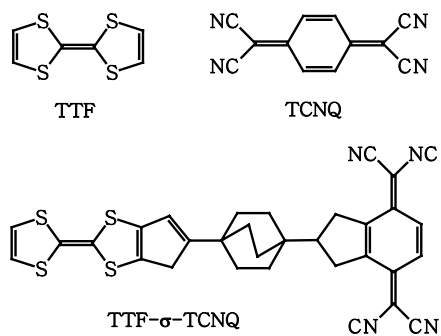

**Figure 2.** Molecular structure of TTF, TCNQ, and TTF-σ-TCNQ.

The true I-V performance of such molecular device may be very complicated due to other contributions to the current. That may include surface terms, direct passage due to inhomogeneous molecular layers, and direct electron tunneling between the electrodes. Several effects were neglected due to the calculation difficulty including the direct energy transfer from donor to acceptor, electrode polarization, and electron correlation. In fact, these effects are very complicated and difficult to understand and control even today more than 30 years later. This is because it is technically very hard to unambiguously construct reliable and addressable single-molecule level metal/molecule/metal junctions, to say nothing of their integration. As a result, the advance of molecular electronics is still far from practical application. However, researchers saw again the hope of using single molecules as data storage media after the invention of the STM. After the concept of TTF-σ-TCNQ, people have synthesized and investigated series of molecule with the structure of donor-σ-acceptor and donor-π-acceptor (Fig. 3) [33,34,35].

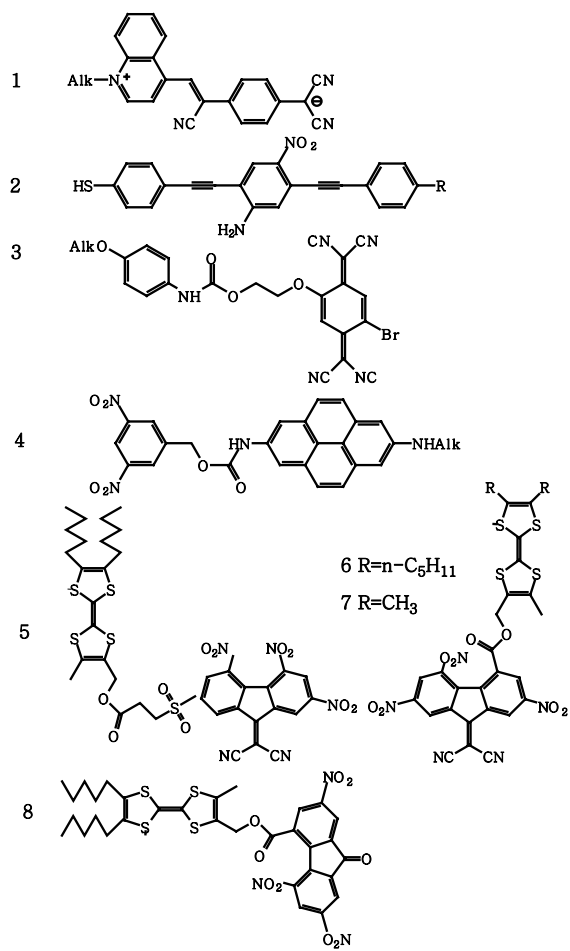

**Figure 3.** Molecular compounds with Donor-π-Acceptor (**1**, **2**) and Donor-σ-Acceptor (**3-8**) structures. Except for **2**, all the molecules are designed with alkane chains for making LB film.

Molecular electronic devices are classified in terms of molecular structure, device type, number of electrode terminals, and state of the molecular media in Table 1. Two-terminal molecular junctions based on monolayers and single-molecules are one of the most important prototypes for nanostorage devices. In this chapter we highlight how to construct addressable, reliable and nanoscale molecular junctions.

**Table 1.** Classification of molecular electronics.

| Material | Device type | Device terminal | Active media |
|---|---|---|---|
| Alkanethiols | Wires | Two-terminal devices | Gas |
| Alkanedithiols | Switches | Three-terminal devices | Solution |
| Conjugated molecules | Diodes | | Thin film |
| Dendrimers | Transistors | | Langmuir-Blodgett film |
| Polymers | Spintronics | | Self assembled monolayer |
| Metal-organic molecules | Cellular automaton | | Pattern of Molecule |
| Organic-inorganic hybrid molecules | Sensors | | |
| Biomolecules | | | |

*2.1 Molecular junctions*

Previous studies on molecular electronics have been carried out suing two-terminal device configurations, i.e., molecular junctions, where the electrical properties of the target molecules are characterized through current-voltage measurements. Various approaches have been proposed for the fabrication of molecular junctions, which can be classified as either solution-based or solid-state junctions. As shown in Fig. 4, for the former case, the target molecules either vapor deposited or self-assembled on a conducting electrode will be measured in electrolyte solution using a counter electrode. This method has the advantages of easy device fabrication and non-destructive characterization, and has therefore been widely used to study different molecular systems, with or without a third terminal. However, a serious disadvantage is that, such molecular junctions cannot be used in practical applications due to the problems of solvent effects, device integration, and addressability.

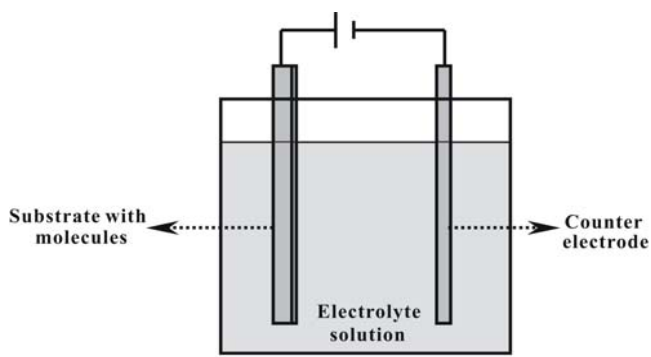

**Figure 4.** Schematic cell structure of electrolyte solution based molecular junction.

In contrast, solid-state molecular junctions have no such problems and thus have been intensively investigated. The fabrication methods were summarized in Table 2 from the aspects of junction size, addressing, and temperature variability. The key findings are:

(1) Although the fabrication of liquid metal droplet and crossed microwire junctions are simple, these microscale molecular junctions are neither stable nor addressable.
(2) The scanning probe microscope (SPM) tip based junctions will be useful in molecular-scale data storage if the "addressing" issue can be solved.
(3) The approaches of crossbar (both lithography and stamp-printing), nanopore and etched-hole plus nanotube are applicable in characterizing the charge transport of molecular systems and probably have potential application in molecular devices with scale between 0.1 and 10 micrometer.
(4) The junctions based on electrodeposited nanowires and mechanically breaking wires are useful in characterizing nanoscale molecular systems. However, due to the "addressing" problem, it will be difficult to use these junctions in practical information devices.
(5) Significantly, the crossed nanowire/tube and nanoparticle bridge based junctions can meet the requirements for nanostorage application from the points of device size, addressability, and reliability. Therefore, we suggest that these technologies are worth further study.

We firstly discuss the development of SPM tip based molecular junctions and then the crossbar molecular junctions will be described. Next, the approaches of nanopore and etching hole are presented, which is followed by the discussion of breaking-wire, electromigration/ electrodeposited nanowire, and nanoparticle-bridge methods. Finally, the metal droplet method and the charge transport mechanisms are summarized in brief.

**Table 2.** Fabrication approaches and characteristics of various solid-state molecular junctions. Here, 'nano' is defined as a scale less than 5 nm.

| Approach | | | Size (d) | Address | Vary T | Nano | Ref. |
|---|---|---|---|---|---|---|---|
| Liquid metal droplet | Hg | | 50~200 μm | No | No | No | [36,37,38] |
| | GaIn | | | | | | [39, 40] |
| SPM tip | STM | STM tip | 1~5 nm | No | No | Yes | [41] |
| | | Nanodot coupled STM | 1~5 nm | Yo | No | Yes | [42] |
| | CAFM | CAFM tip | 5~50 nm | No | No | No | [43] |
| | | Nanodot coupled CAFM | 1~5 nm | Yo | No | Yes | [44] |
| Crossbar | Magnetically controlled crossed-wire | | 0.5~5 μm | No | No | No | [45] |
| | Lithography crossbar | | 2~50 μm | Yes | Yes | No | [46,47] |
| | Stamp-printing crossbar | | | | | | [48,49,50] |
| | Crossed nanowire/tube | | 1~5 nm | Yes | Yes | Yes | [51] |
| Etching hole plus nanotubes | | | 2~10 μm | Yes | Yes | No | [52] |
| Nanopore | | | 30~50 nm | Yes | Yes | No | [53,54,55] |
| Electromigration | | | 1~70 nm | Yes | Yes | Unknown | [56,57] |
| Electrodeposited nanowire | | | 1~50 nm | Yes | Yes | Unknown | [58,59,60, 61,62,63] |
| Mechanically breaking wire | | | 1~50 nm | No | No | Unknown | [64] |
| Nanoparticle bridge | | | 1~5 nm | Yes | Yes | Yes | [65] |

SPM = Scanning probe microscope  
STM = Scanning tunneling microscope  
CAFM = Conducting atomic force microscope  

d = Diameter  
Vary T = Variable Temperature  
SAMs = Self-assembled monolayers  

A. STM tip based molecular junction

One way to achieve nanoscale molecular junctions is to use STM tip-assisted methods. STM can not only image the surfaces of conducting or semiconducting materials with atomic resolution but can also provide the information on their electronic information via I-V measurements. There are two main device configurations for STM tip based molecular junction (Fig. 5). One configuration is tip/gap/molecule/substrate [41], the other is tip/gap/nanoparticle/molecule/substrate [42]. The STM tips can be mechanically cut PtIr and Au or chemically etched tungsten wire. The substrate can be surfaces of single-crystal metal or highly oriented pyrolytic graphite (HOPG). The molecular layer may be a SAM, LB layer, or even a thin film. Obviously, it will be difficult to use organic thin films as the active medium in nanostorage due to the large roughness and poor uniformity [66].

To form a molecular junction, the molecule of interest can be self-assembled onto Au (111) surface via thiol-Au bonds and the STM tip is then used to probe the electronic properties of the molecules. Single-molecule level junctions can be obtained with using a 'nonconducting' molecular matrix technique, i.e., by inserting the molecule of interest into a relatively insulating molecular monolayer (e.g. dodecanethiol) [41]. However, owing to the thermal drift of the STM instrument, it is extremely difficult to measure the electrical properties of the same molecule for a long time. Therefore, such molecular junctions are not considered to be addressable and no data of I-V temperature dependence is available. Moreover, there exists a tunneling gap between the tip and the molecular layer. This tunneling gap can result in some non-molecule-specific device performances such as electrical rectification.

Addressable single-molecule junctions have been achieved using the nanoparticle coupled method, though there is no way to eliminate the tunneling gap between the tip and the particle. In this case, nanoscale metal clusters need to be covalently linked with dithiol-ended molecules. For example, Au clusters had been rigidly attached to a dithiolated molecular monolayer assembled on an Au (111) substrate [42]. Moreover, a reliable approach to measure single-molecule properties is to take the advantages of nanoparticle coupled method and the insulating molecular matrix techniques [67,68]. The nanoparticles and thus the dithiolated molecule of interest can be reproducibly addressed using a STM tip. But this does not guarantee the addressing and reproducibility of the junction because the operation of the STM is extremely sensitive to the quality of the tip. If the STM tip adsorbs some contamination or even the molecule of interest, the measured electronic properties may be drastically affected and the consequent STM image might change a lot too. In an extreme case, if the tip is damaged by collision with the substrate, the target molecule will get lost. Obviously, such molecular junctions are not suitable for practical applications.

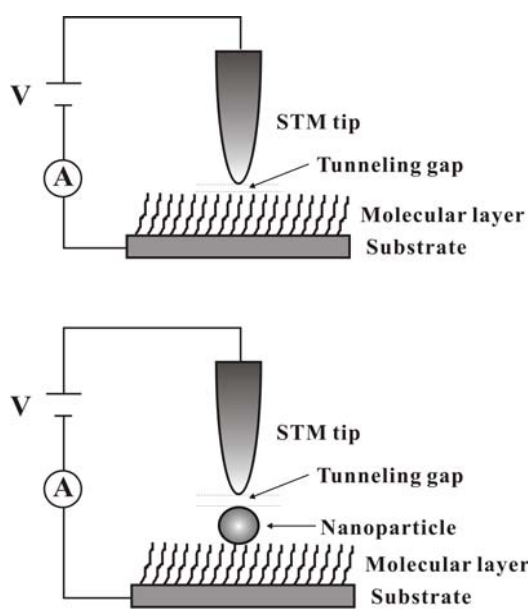

**Figure 5.** Schematic structure of STM tip based molecular junctions without/with nanoparticle coupled. Note that there exists a tunneling gap between the tip and the molecular layer or nanoparticle. The tip can be mechanically cut PtIr, Au, or chemically etched tungsten.

STM has also been used to construct molecular break junctions. Tao and co-workers measured the conductance of single molecules using this method [69,70,71]. Briefly, molecular junctions were created by repeatedly moving an Au STM tip into and out of contact with Au substrate in solution containing the target molecule (~ 1 mM). The breaking process was controlled by the feedback loop of the STM. Using this statistical method, the conductances of various molecular wires were calculated from histograms of the conductance curves. However, there are some problems associated with this approach. First, these break junctions are not addressable. Second, it is not known if double-layer molecular junctions with the structure of tip/SAM/gap/SAM/substrate were formed. Because the experiments were conducted in solution with thiolated molecules, SAMs can form on the surfaces of both the substrate and the Au tip. Third, it is difficult to evaluate the effect of the chemical solvent on the device performance. Therefore, a modified STM tip based break junction method has been developed to deal with some of these issues [72]. The main modifications were: 1) The molecular SAM was rinsed and blow-dried with nitrogen before any I-V measurements, while the molecule of interest was formed onto Au (111) surface from 1 mM ethanol solution. 2) The break junctions were measured under ultra-high vacuum conditions. In this case, the solvent effect and possibility of double-layer junction can be excluded.

Also, STM tip based atomic and molecular manipulation has been suggested for application in nanostorage too [73,74], but the problem of bit addressing still limits their potential applications.

B. CAFM tip based molecular junction

Conducting atomic force microscopy (CAFM) is also used to study the electrical property of molecular wires. There are two approaches to achieve CAFM tip based molecular junction (Fig. 6). One is to directly place the tip in contact with the molecular layer (e.g. SAM, LB, or thin film) under a small load. Then the I-V data is collected at various applied forces and sites [43]. Unlike STM methods, there is no tunneling gap in the CAFM tip based molecular junctions, which may decrease the difficulty to analyze the I-V data. The tradeoff is the big increase in the junction size as a result of the large radius of the CAFM tip (30~50 nm) as compared to that of the STM tip (few angstrom). It was estimated that a CAFM tip based junction contains at least 50 molecules. Au nanoparticles were used to construct single-molecule junction with a combination of CAFM tip and molecular matrix techniques [44]. The junction size depends on both the nanoparticle diameter and the load of the CAFM tip. The molecular junctions based on CAFM tip are unstable and difficult to integrate. Moreover, it is impossible to investigate the temperature dependence of the molecular conductance using this approach, while temperature dependence data is critical for understanding of the charge transport mechanism.

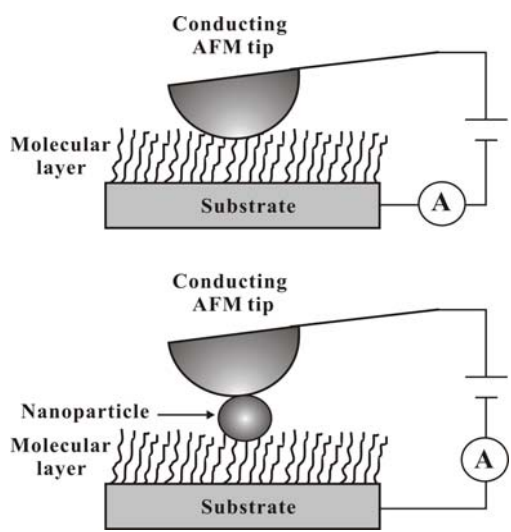

**Figure 6.** Schematic representation of CAFM tip based molecular junctions without and with nanoparticle coupled, respectively. To form a molecular junction, a small force (~2 nN) has to be applied between the CAFM tip and the molecular layer/nanoparticle.

C. Crossbar molecular junction

Molecular junction can be formed by two crossed metal wires, in which one wire is assembled with the target molecule. The electrical property of the molecules is investigated through I-V measurements under different conditions, such as variations in temperature, pressure, and/or light irradiation. As demonstrated in Fig. 7, the crossbar method may also be used to build molecular junction array, which can be assembled and programmed to perform fast, powerful, and reliable functions like data memory and communication with defect-tolerant characteristics [75].

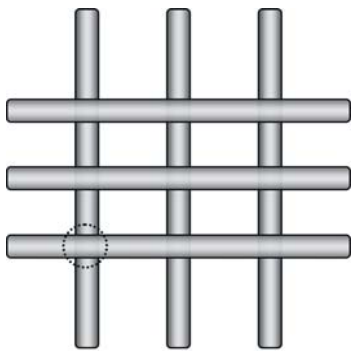

**Figure 7.** Schematic representation of a 3 × 3 crossbar junction arrays with one outlined by the dashed circle.

As shown in Table 2, there are four approaches to make crossbar junctions: magnetically controlled crossed-wires, photolithography crossbars, stamp-printed crossbars, and crossed nanowire/tubes. The simplest junction is based on two crossed gold wires (10 ~ 50 μm in diameter) with one of them modified with a SAM of the target molecule [45,76,77]. The wire spacing is controlled by Lorentz force: the dc current in one wire deflects it in a magnetic field. The deflection current is slowly increased to bring the wires into gently contact and a junction is thus formed. Three different molecules have been tested using this method: 1,12-dodecanethiol (C12), oligo(phenylene ethynelene) (OPE) and oligo(phenylene vinylene) (OPV) (Fig. 8). The molecular conductance follows the trend of OPV> OPE > C12. However, these molecular junctions have some disadvantages in practical application because they are neither addressable nor stable.

Addressable crossbar junction arrays with LB monolayer of **$C_{20}$**, **DB**, or **R** have been fabricated using photolithographic technique (Fig. 8) [46,47]. The bottom electrode bars (1~10 μm-wide) and bonding pads (>100 μm × 100 μm) were formed by photolithography on silicon substrate with thermal oxide layer (100 nm-thick). The electrode surface was cleaned by organic solvent and oxygen reactive ion etching (RIE), molecular monolayer was then self-assembled onto it through Au-thiol bond or LB technique. For the junction with LB layer, the bottom electrode can be a 100-nm-thick Al or Pt film [46,47]. On the other hand, an Au film with 5-nm adhesion layer (Cr or Ti) is normally used in junction with self-assembled monolayer. To complete the fabrication, top electrode bars (5~20 μm-wide) and bonding pads were evaporated through a shadow mask on the top of the molecular layer. To prevent metal filament formation, a 5-nm-thick Ti reactive layer was deposited before the deposition of thick (50 ~100 nm-thick) Al, Pt, or Au film.

The junctions were characterized through I-V measurements. Reversible electrical switching was observed from junctions with the structure: Al/LB monolayer of **R**/Ti/Al [46], which was ascribed to an intrinsic property of the redox molecule **R**. Latterly, it was later found that similar switching also occurred in crossbar junctions with non redox-active molecules (e.g., **$C_{20}$**) [47]. It suggests that the metal/molecule interface may play a significant role in such switching performance. On the other hand, there are some problems with this crossbar method: 1) the deposition of the Ti reactive layer may cause thermal damage to the organic monolayer; 2) the good device rate is very low because of pinhole-induced short circuit. Especially, if we consider the big size of the junction (>5 μm$^2$); 3) it is difficult to scale down the junction size to sub-5 nm dimension due to limitations in scaling down the shadow mask.

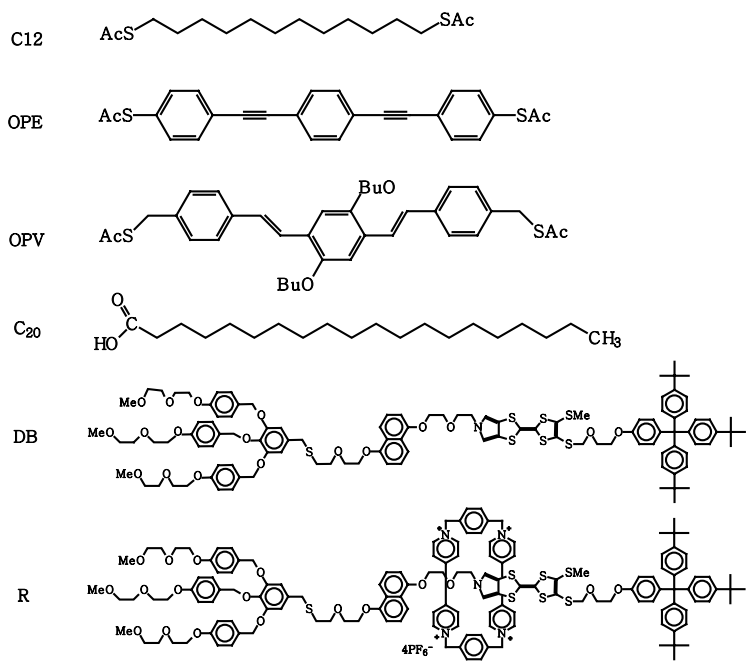

**Figure 8.** Chemical structure of molecules studied using magnetically controlled crossed Au wire (C12, OPE and OPV) and photolithography crossbar junctions (C20, DB and R) [45,46,47].

An alternative method to fabricate the top electrode of the crossbar junction is solid-state metal transfer techniques. For instance, a poly(dimethylsiloxane) (PDMS) stamp-printed crossbar method has been developed to make addressable molecular junctions with optoelectronic switching functions [48]. Fig. 9 shows the fabrication procedure. Briefly, the crossbar sample was made with the following steps:

(1) Fabricate the bottom Au electrode pattern by photolithography. First, the Si (100) substrate with 250 nm-thick $SiO_2$ on both sides was cleaned in hot "piranha solution" (a 4:1 v/v mixture of concentrated $H_2SO_4$ and 30% $H_2O_2$) for 10 min followed by rinsing with deionized water, acetone, and ethanol. *Caution: Piranha solution is a highly oxidizing solution that may react violently with organic materials.* Then, the gold pattern is defined on the substrate by photolithography and vacuum vapor deposition. The mask design and a scanning electron microscope (SEM) image of several junctions are shown in Fig. 10. The photolithography parameters are given in Table 3.

(2) SAM was formed on the gold bar surface by immersing the substrate in a 1 mM molecular solution in pure enthanol or tetrahydrofuran (THF) for 24 hrs. The substrate was thoroughly washed in pure organic solvent and blow-dried just before stamp printing. To reduce the number of monolayer defects, the SAMs of conjugated molecules were all made from a mixed solution with decanethiol in a mole ratio of 1 to 1.

(3) The PDMS stamp was fabricated via casting and curing a prepolymer against negative photoresist model [78].

(4) Deposit a 20 nm Au film onto the stamp (cooled by liquid nitrogen) using vacuum evaporation.

(5) Under an optical microscope, the Au-coated stamp was brought into contact with the substrate Au bars in perpendicular direction.

(6) This stamp/substrate set was immediately loaded into a vacuum chamber and cooled down to 100 K for 10 minutes. After warming up, the stamp was carefully peeled off and the crossbar junctions were obtained with good device rate of 15%.

(7) Last, wiring the sample and load it into a vacuum chamber and measure the I-V characteristics under different conditions.

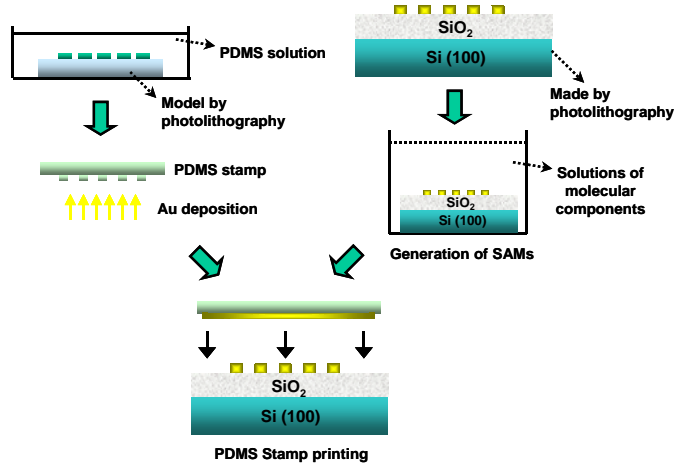

**Figure 9.** Schematic diagrams showing the procedure of molecular crossbar fabrication using PDMS stamp-printing techniques [48].

**Table 3.** Photolithography parameters for making PDMS stamp-printing crossbar junctions. Note that a temperature-increasing step is needed in the baking processes of negative photoresist.

| Process | Photoresist patterns | |
|---|---|---|
| | For Au bars on Si (100) wafer | For PDMS model |
| Photoresist | Positive resist S1813 | Negative resist SU-8 |
| Spin-coat | 3.5 krpm, 60 s | 3 krpm, 80 s |
| Soft-bake | 115 °C, 60 s | 65 °C, 120 s → 95 °C, 300 s |
| Cooling | 180 s | 180 s |
| Exposure | 60 s | 85 s |
| Hard-bake | 115 °C, 180 s | 65 °C, 60 s → 95 °C, 120 s |
| Cooling | 180 s | 180 s |
| Develop | 45 s | 90 s |
| Clean | Thoroughly rinse with deionized water flow and blow-dry with nitrogen | |

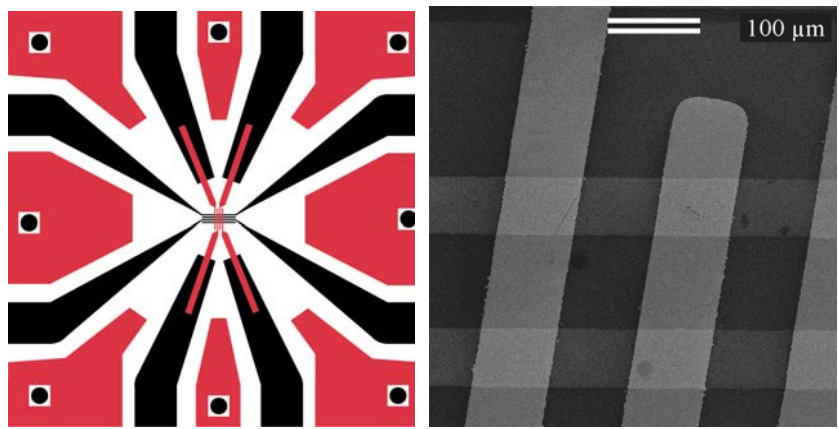

**Figure 10.** (Left) Schematic diagram of mask design for PDMS stamp-printing crossbar experiment. The black color stands for the bottom electrode pattern defined using optical lithography, while the red one represents the top pattern printed by PDMS stamp. Ohmic contacts between the printed Au bars and their bonding pads were ensured by a droplet of GaIn liquid metal through a sharp tungsten tip controlled by micrometer. (Right) A SEM image shows several crossbar junctions [48].

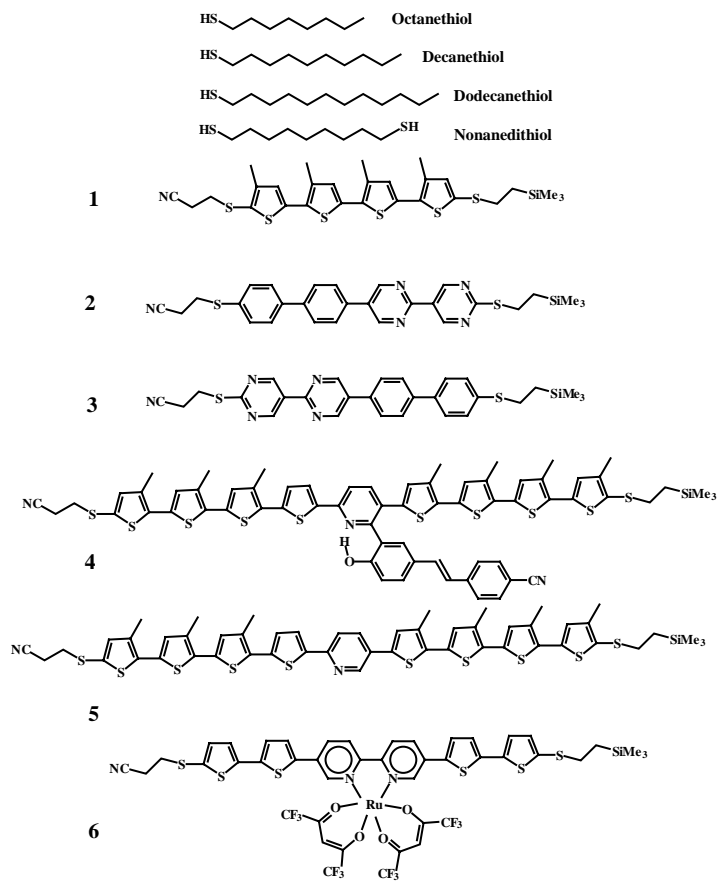

**Figure 11.** Chemical structure of the molecules investigated using the testbed of PDMS stamp-printing crossbar junction [48].

Fig. 11 shows the chemical structure of the molecular wires studied using PDMS stamp-printed crossbar approach. The molecules include not only alkanethiols (from $C_8$ to $C_{12}$) and alkanedithiols, but also conjugated wires with different chain lengths, molecular dipoles and/or metal cations. The I-V measurements were conducted in high vacuum condition at temperature from 95 to 300 K. Asymmetric I-V curves were observed for all the molecular junctions studied, which strongly depend on the temperature. The electrical performance of the crossbar junctions is found to be extremely sensitive to optical illumination despite wavelength variation (254 ~ 700 nm). Fig. 12 presents the optoelectronic switching of a nonanedithiol junction under stimulus of a fluorescent light signal. More interestingly, the conductance of the molecular junctions can be reversibly tuned via varying the light intensity [48], which may make them serve as a prototype for optical gated molecular transistors. It is indicated that the switching is due to light induced change of the tunneling parameters at the top PDMS-printed Au/molecule interface. The size of the molecular crossbar junctions can be further scaled down to sub-50 nm using hard-stamp imprinting methods [79,80].

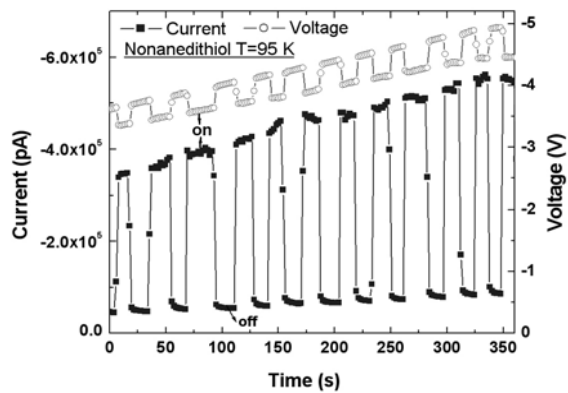

**Figure 12.** Reversible optoelectronic switching of 1,1-nonanedithiol crossbar junction at the temperature of 95 K. The optical signal is fluorescent light with wavelength of 400~700 nm. [48].

Crossbar junctions comprising a LB monolayer of [2]rotaxanes **R** (Fig. 8) have been built using nanoimprint lithography (NIL) technique with device dimensions of 40 nm × 40 nm [49]. The imprint mold was fabricated into 100-nm thick $SiO_2$ layer grown on a silicon substrate by e-beam lithography and fluorine-based reactive-ion etching (RIE) [80]. The oxide surface was patterned and etched to the form of 40-nm-wide nanowires connected by micrometer scale wires on each end to hundred-micrometer scale bonding pads. To make the bottom electrodes and bonding pads, 100-nm thick PMMA (polymethylmethacrylate) film is spin-coated onto a silicon substrate with 100-nm thermally grown oxide layer. The mold and PMMA film are heated to 150~200 °C and then the mold is pressed against the PMMA substrate until the temperature drops below 105 °C. After removing the mold, oxygen RIE is used to etch the residual PMMA at the bottom of the trenches to expose the $SiO_2$ surface. Next, 10-nm-thick Pt nanowires (with 5-nm-thick Ti adhesion layer) are fabricated using vapor deposition and lift-off techniques. An LB monolayer of **R** is then deposited over the entire substrate with the bottom electrodes, which is followed by a blanket evaporation of 7.5-nm Ti protective layer onto the monolayer. Finally, the top perpendicular nanowire patterns (10nm Pt/5 nm Ti) are prepared using the same imprint process. Electrical switching behavior was observed from the I-V measurements under ambient conditions, which has been proved to be independent of both the molecular structure and other molecule-specific properties [47]. Therefore, the switching may be related with the metal/molecule interfaces.

There are two problems with this imprint method. One is with the thermal deposition of the Ti protective layer. Though Ti layer is very reactive with organic molecules, the thermally evaporated Ti clusters may still penetrate into the monolayer. This will consequently result in metallic filaments formation owing to the defects and pinholes of the monolayer [47]. The other problem is with the imprint fabrication of the top electrode wires. Because undesired contaminations and/or damages may be made to organic monolayer in the procedures involving the PMMA spin-coating, the high temperature contact-imprinting, the oxygen RIE, and the acetone lift-off.

To circumvent these problems, an improved imprint approach has been designed to fabricate molecular crossbar junctions [50] (Fig. 13). First, a stack of 220 nm silicon nitride ($Si_3N_4$), 250 nm $SiO_2$, and another 220 nm $Si_3N_4$ films are grown on a silicon wafer using low-pressure CVD (chemical vapor deposition) and plasma enhanced CVD, respectively. A layer of imprint polymer is then spin-coated and patterned using a NIL hard-stamping mold. Using this patterned polymer as a mask, $CHF_3$ reactive ion etching is used to etch down to the oxide layer and thus produce a patterned trench. The residual polymer and oxide are then stripped with acetone and etched to the bottom nitride layer with diluted HF acid, respectively. Note that under-etching of the top nitride layer by about 400 nm is necessary to obtain good device. The I-V curves of 1-octadecanethiol SAM show an electrical rectification with a value of about 2 at ± 0.8 V, which is claimed to be due to the extraordinarily different metal/molecule interfaces: one is strong Au-thiol chemical bond and the other is weak physisorption. However, there is no temperature dependence I-V data for understanding the rectification mechanism.

One problem of this imprint method is that, although the substrate is cooled using a liquid nitrogen stage, the deposition of the top Au electrodes may induces thermal damage to the organic monolayer and consequently results in nanofilaments formation and junction short-circuit. This will not only lead to a very low good device-manufacturing rate, but also result in artificial effects in the devices [81,82]. More critically, it will be very difficult to use this method to obtain crossbar junctions with dimensions scale down to sub-10 nm.

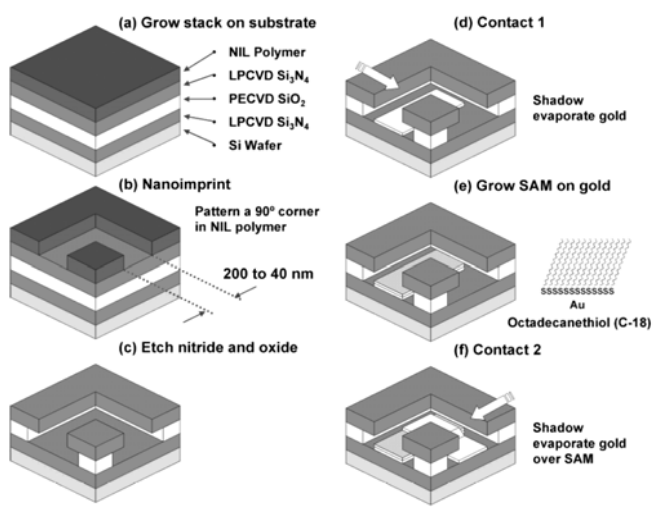

**Figure 13.** Schematic fabrication process for nanoimprint lithography crossbar junction. (a) Initial stack on silicon substrate. (b) Nanoimprint lithography is used to make pattern on imprint polymer. (c) Form a trench using Reactive Ion Etching and then use HF acid to selectively etch the $SiO_2$. (d) Form the first electrode contact by shadow-mask deposition of Au in one trench. (e) Make SAM of 1-octadecanethiol onto the Au surface. (f) The second Au contact is deposited over the SAM down the other trench. [After ref. 50]

The nanowire/nanotube crossbar approach provides a promising way to fabricate addressable and reliable molecular junctions. It was demonstrated that molecular junctions can be made through a combination of carbon nanotube and shadow-mask deposited Al electrodes (with Ti adhesion layer) [83]. But, the micro-scale dimensions and the thermal deposition of the top electrode make this method unfavorable for application in nanostorage. Nanoscale molecular crossbar junctions have been fabricated using nanowire assisted e-beam lithography (EBL) technique [51]. Briefly, large bonding pads and electrodes are firstly fabricated on SiO2/Si substrate by photolithography. Then, EBL is carried out to write hundred nanometer scale electrodes with connection to the large ones. The next step is to self-assemble the target molecules onto the nanoelectrodes in solution. For the top electrodes, conductive nanowires with diameter of 10~15 nm are spun onto the sample surface from a suspension of pre-made Pd nanowires in dichloroethane. Electrical switching and memory performances are observed in the crossbar junctions with molecule **1**, **2,** and **3**, respectively, while no similar behaviors are found in junctions with **4** or **5** (Fig. 14). It was suggested that the electron-retracting groups (i.e., nitro or pyridine) play a key role in the observed memory performance.

A concern is the electrical contact between the nanowire and the EBL defined electrode with SAM. To eliminate the formation of double junctions, an AFM tip has been used to scratch the SAM off the electrode and expose the gold surface. However, it is doubtable if the loading force is big enough to scratch off molecules chemically bonded on Au surface. Even if there is no SAM between the nanowire and the scratched electrode, the contact is not guaranteed to be Ohmic. Because the nanowires are spin-coated onto the substrate from solution, the interaction between the nanowire and the electrode is only weak physisorption. Another concern is how to make sure the nanowire suspension is free from undesired residues that may affect the good junction manufacturing rate and device performance to a great extent.

Recently, an experiment revealed that no such electrical switching observed in similar molecular junction fabricated using a different method (Fig. 14 (**6** ~ **11**)) [52]. This observation clearly indicates that molecular junctions with different size, electrode material, and/or fabrication method may show tremendously different device performance. To make sure that all performances result from molecule-specific properties, rather than other factors, we need to examine the target molecule using different fabrication approaches. Single-molecule level junction is helpful to eliminate most non-molecular factors. As shown in Fig. 15, this can be realized using a nanowire and/or nanotube crossbar configuration. The target molecule can be self-assembled onto a sub-10 nm Au wire connected with large bonding pads and then another bare Au nanowire is perpendicularly placed on top. In this case, the junction size may be so small that it contains just one or two molecule(s). Experimentally, it has been shown that addressable crossed nanowire/tube arrays can be successfully constructed using different methods [11,84,85,86,87,88]. Theoretical simulations were performed to assess the prospective performance of a 16 Kbit crossed nanowire nanomemory system [89]. The results suggested that such nanomemory system can operate at a density of no less than $10^{11}$ bits/cm$^2$. To solve the problem of double junctions, the molecule of interest can be controllably deposited onto the desired locations of the nanowire/tube using techniques such as dip-pen nanolithography and stamp-printing method [90,91].

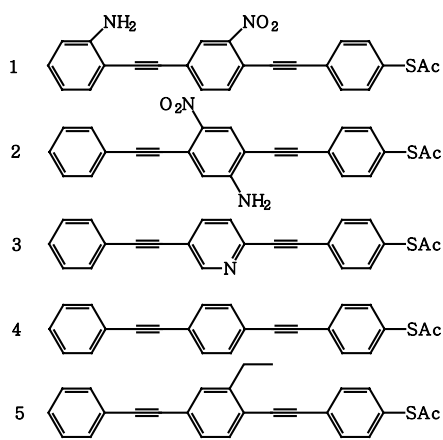

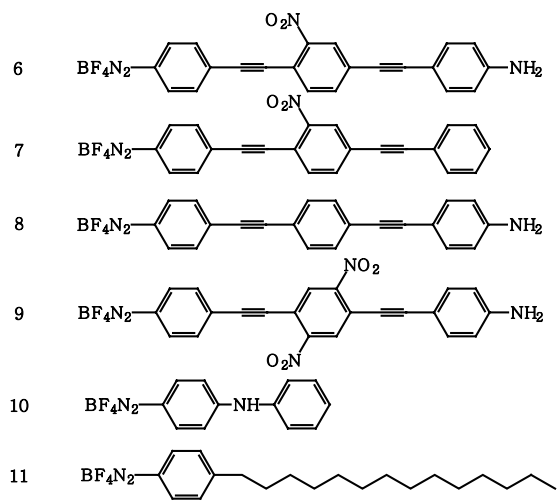

**Figure 14.** Chemical structure of molecules tested using approaches of crossed nanowire (**1-5**) [51] and etching hole plus nanotube (**6-11**) [52].

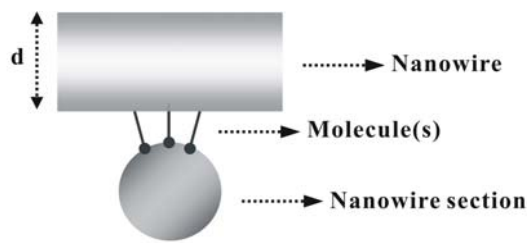

**Figure 15.** Schematic diagram of crossed nanowire molecular junction. The nanowire diameter (d) should be less than 10 nm.

D. Etching hole plus nanotube and Nanopore molecular junctions

A silicon/molecule/nanotube approach has been developed as a testbed for fabrication of metal-free molecular junctions, i.e., etching hole plus nanotube method [52]. This approach is designed to circumvent the problems of thermal damage and metal filament formation in the monolayer induced by thermal deposition of the top electrodes. The junctions are fabricated on highly doped n-type Si(100) substrate (<0.005 Ω) with 200 nm thermal oxide layer. First, Au bonding pads (200 nm-thick with 20-nm Ti adhesion layer) are made on the substrate by photolithography. A second photolithography step is followed to form circular well patterns (5 to 20 μm in diameter) between the pads. The wells are then wet etched down to the doped Si layer in a buffered oxide etch and this affords a H-passivated Si surface. A molecular monolayer is thereafter self-assembled in the well via direct Si-arylcarbon bond [92]. Fig. 14 (**6 ~ 11**) shows the chemical structure of the molecules studied. A purified carbon nanotube mat is deposited from a suspension in chloroform onto the top of the well region to make electrical connection to the neighboring Au pads. A 200-nm Au film is sputter-coated on the backside of the silicon substrate to serve as the bottom electrode. The device I-V characteristics are finally collected at different temperatures in a cryostation.

Although this metal-free method can avoid the thermal damage to the molecular monolayer and metal filament formation in the monolayer, there are some other problems with the junction. One question is about the electrical contacts between the Au pads and the carbon nanotubes. Are they Ohmic or non-Ohmic contacts? What is the effect of these contacts on the electrical performance of the molecular junctions? The other problem is that the actual 'effective' junction size is unknown owing to the uncertainty of the contacting area between the nanotube mat and the molecular monolayer. This is true, especially, if we take into consideration the comparison of the ultrathin monolayer with the roughness of nanotube mat on top.

A nanopore approach has been designed to overcome these problems [54]. As shown in Fig. 16, arrays of 30~45 nm-diameter pores are opened in suspended SiN membrane deposited on Si wafer using e-beam lithography and plasma etching [53,54,55,81]. Gold film is evaporated from the bottom side to fill the bowl shaped pores. Then, the sample is immediately immersed into millimolar molecular solution in pure ethanol for 24 hrs under the protection of inert gas. After being thoroughly washed with ethanol and blown dry with nitrogen, the sample is transferred into a vacuum chamber to deposit the top electrode, which can be either a Ti/Au (4 nm/80 nm) bilayer [53] or a pure Au (200 nm) film [55]. Meanwhile, the sample has to be kept at liquid nitrogen temperature to minimize the thermal damage to the SAM and prevent metal filament formation. Different molecules have been tested using this method, which includes 4-thioacetybiphenyl (Fig. 16), 1-octanethiol, 1-decanethiol, and 1-hexadecanethiol [53,55].

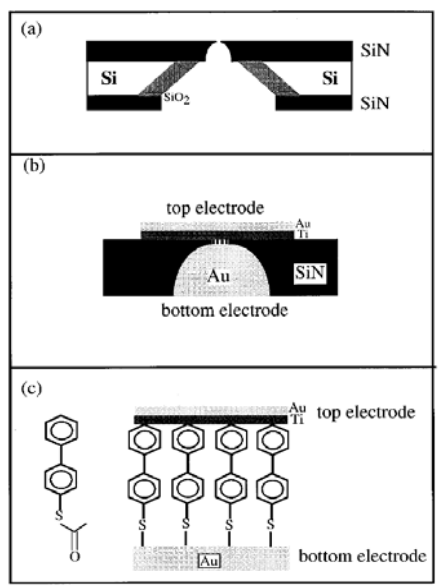

**Figure 16.** Schematic diagrams of the fabrication process of nanopore junction. (a) Cross section of a nanopore (30 nm-diameter) etched in suspended SiN membrane. (b) Molecular sandwich structure of Au-Ti/molecule/Au. (c) Chemical structure of 4-thioacetylbiphenyl and detailed structure of the junction. [After ref. 53]

Molecule **2** in Fig. 14 has also been studied using the nanopore method, in which electrical switching and negative differential resistance behaviors are observed [81]. However, it was latter found that such performance are probably due to the metal/molecule interfaces and/or metal nanofilament formation [82]. This result reminds us again that the device performance of a molecular junction strongly depends on the local environment including electrode/molecule interface, fabrication method, temperature, humidity, optical illumination, and intermolecular interaction. Before these problems can be solved, it will be very difficult to discuss which, or what kind of, molecules are more suitable for nanostorage application. Actually, it is for this reason that we limit this review to the issue of nanostorage device fabrication rather than to the molecular material and device mechanism.

E. Mechanically breaking wire, electromigration-induced breaking wire, and electrochemically deposited nanowire junctions

Charge transport through molecular junction of benzene-1,4,-dithiol has been investigated using mechanically breaking wire approach in 1997 [64]. The experiment is conducted at room temperature with a notched micrometer-scale gold wire immersed in 1 mM molecular solution in THF. The wire is glued onto a flexible substrate held by two mechanical supports. An atomically sharp tunneling gap can be established after the notched wire is fractured by a counter pizeo element. The electrical measurements are conducted after the THF was evaporated under argon atmosphere. Due to the thermal gradients induced by THF evaporation, the tunneling gap of the break junction will be disturbed. The tips have to be withdrawn and then carefully moved together to an onset of conductance before any I-V measurements (Fig. 17). This withdrawing/returning process can be repeated many times to test the reproducibility of the experiment. The break junction is neither stable nor addressable. As a result, it is impossible to measure the effects of temperature, pressure, or optical illumination.

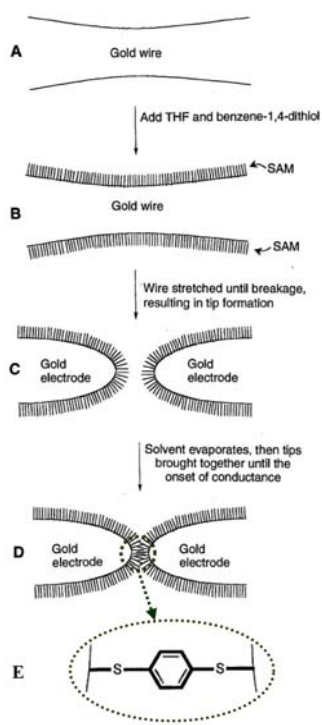

**Figure 17.** Schematic process of break junction formation. (A) The thin Au wire with an etched neck. (B) SAMs formed on the wire surface after adding the molecular solution. (C) Mechanical breakage of the wire and SAMs formed onto the surfaces of the Au tips. (D) After evaporation of the THF, the tips were carefully brought together to the onset of conductance to collect I-V data. (E) A benzene-1,4,-dithiol is schematically bridged between two gold electrodes. The processes of (C) and (D) can be repeated for many times to test the reproducibility of the experiment. [After ref. 64]

As shown in Fig. 18, there are some alternative ways to solve the above problems. One method is to fabricate molecular junctions by passing a large electrical current through a gold nanowire pattern defined by e-beam lithography. The current flow causes the electromigration of gold atoms and consequently the breakage of the nanowire at the narrowest site with a separation of about 1 nm [56]. This is known as the electromigration breaking wire method. It has been used to measure the conductance of CdSe nanocrystals [56], C60 [57], C70, and several other molecules shown in Fig. 19 [93,94]. The key issue for a successful experiment is to control the gold nanowire thickness to sub-20 nm. This can be realized by creating a 200-nm-wide PMMA resist bridge suspended 400 nm above the substrate at the neck part of the EBL defined nanowire (indicated by the dashed circle in Fig.18). EBL on a PMMA/P(MMA-MAA) bilayer resist is employed to produce a bridge through undercutting effect, i.e., the different developing rate of the two kinds of resist layer. Then ±15° angle (respect to substrate normal) evaporations are conducted to generate a gold nanowire with a thickness less than 20 nm. Finally, a thick metallic film is vapor deposited straight down through the bridge to make a good connection between the nanowire and the bonding pads. The advantage of this approach is it does not use any organic solution in the process of the nanoscale gap formation, although the resist bridge fabrication and angle deposition make it complicated to some extent. However, as a big challenge, the junction is difficult to scale down to sub-5 nm widths. Another challenge is with the integration of individual junctions into functional systems.

The other method is to use electrochemical deposition to fabricate nanowire junctions. A pair of planar electrodes with a gap between 50 and 2000 nm is firstly made using e-beam lithography. Electrochemical deposition is then applied to shrink the gap to a sub-5 nm scale by applying a bias voltage between the electrodes in electrolyte solution. The gap distance is controlled through *in situ* monitoring the resistance between the electrodes [62]. The molecule of interest is then assembled into the nanoscale gap from molecular solution to measure the I-V characteristics. Different nanowire (such as Au, Cu, and Pt) junctions can be made using this technique [59,60,61,62,63]. It is simpler than that of the electromigration method, since there are no resist bridge and angle depositions needed. As a tradeoff, the junction has to be fabricated in the presence of solvent, which may introduce some problems to the device performance.

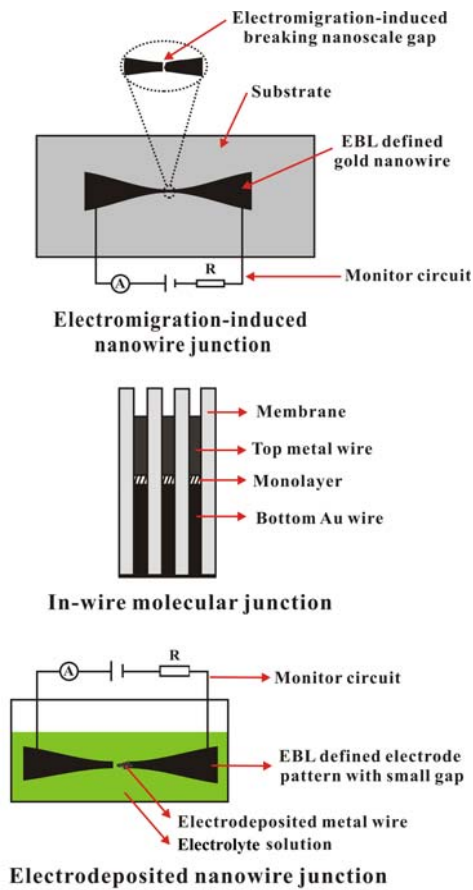

**Figure 18.** Schematic setups for fabrication of nanowire molecular junctions. For the electromigration method, a gold wire pattern with a narrow neck is firstly defined by e-beam lithography and then a high current flow is used to induce a nanoscale gap (~1 nm) (no solution needed in this step) [56]. For the case of in-wire junction, a polycarbonate membrane with 70 nm-diameter pores is employed as template to grow the bottom half Au nanowire, molecular monolayer, and the top half nanowire, respectively. The nanowires are released by dissolving the membrane before I-V characterizations [58]. For electrodeposited nanowire junction, a gold electrode pattern with a small gap (100 nm to 2 μm) is firstly fabricated on a substrate. The gap is then decreased through electrochemical deposition of metal atoms onto one of the electrode in electrolyte solution. The gap distance is controlled by monitoring the junction resistance [59,60,62].

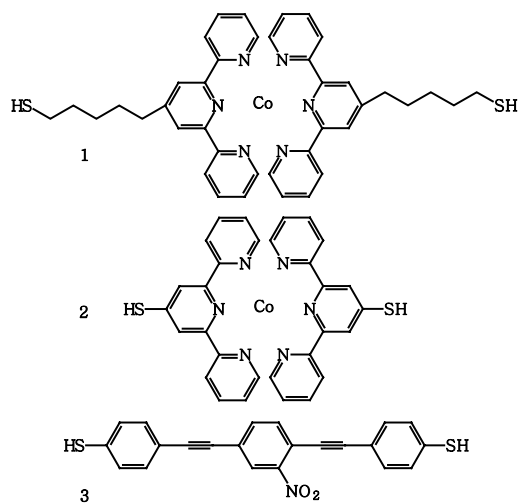

**Figure 19.** Chemical structure of the molecules characterized using electromigration nanowire junction approach.

In-wire junction method has also been used to fabricate molecular junctions [58,95]. A track etched polycarbonate membrane with nanopores (40 ~70 nm in diameter) is used as a template to make the junction. First, a 100 nm-thick Au film is vapor deposited on one side of the membrane to serve as the cathode. Then Au is plated at a constant potential into the nanopores from electroplating solution using a three-electrode electrochemical cell. The molecule of interest is then self-assembled onto the surface of the Au nanowires inside the pores from 0.5 ~ 1.0 mM molecular solution in pure ethanol under inert gas atmosphere for 24 hrs. The monolayer is then capped with Pd or Ag seeding nanoparticles through electroless plating. The top portion of the pores is finally filled with the second half Au nanowires using electrochemical deposition again. The nanowires are released by dissolving the membrane. For I-V measurements, individual nanowires are positioned between EBL defined electrode patterns using electrofluidic assembly.

Although molecular junction can be made using this nondestructive in-wire electrodeposition method, there are some important issues that need to be addressed. First of all, as compared to the other two methods, it is not easy to assemble these nanowires on a large-scale between the electrode patterns pre-defined by EBL. Second, the electrical contacts between the nanowires and the EBL defined electrodes are not well controlled. These wires just loosely reside on the electrode surface and they are not necessarily Ohmic contacts. This may result in extra-high contacting resistance and consequently low power efficiency. Third, it is difficult to scale the junction dimensions down to sub-10 nm due to limitations imposed by the template pore size. Another big challenge is how to eliminate the contamination induced by the presence of electroplating solution during the nanowire preparation and electrofluidic assembly. Moreover, it is also doubt about the monolayer quality, because it is difficult to thoroughly wash the molecular monolayer inside the pores.

However, hope emerges from these works. As shown in Fig. 20, if we can assure extra-clean solutions and solve the problems associated with the molecular assembly and electrode contact, then by utilizing a combination of techniques [96,97,98,99], we may develop nanoparticle-coupled multiple molecular junctions in a single nanowire. If this design can be experimentally realized, it might be an important step toward the generation of bio-molecule-nanostructure hybrid nanostorage systems.

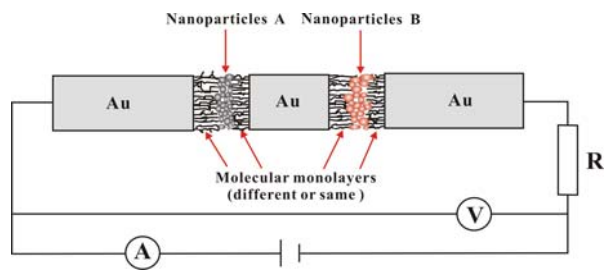

**Figure 20.** Schematic setup of a nanoparticle-coupled multiple molecular junctions fabricated within a singe nanowire.

F. Nanoparticle bridged molecular junctions

Conductance of single conjugated molecule has been measured using nanoparticle bridged molecular junctions [65]. The basic concept of this method is schematically shown in Fig. 21 (top). Colloidal Au nanoparticles with relatively large diameters (10 ~ 50 nm) are firstly prepared using a modified chemical method [100]. The next step is to synthesize dimers with the structure of two gold particles connected by a dithiolated molecule, which is realized by mixing a solution of dithiolated molecule with the gold colloid but keeping the molecule: particle molar ratio below 1:10. The dimeric Au particles are then separated from other particles and oligomers by centrifugation. Then, electrostatic trapping is performed by putting a drop of the dimer solution onto EBL defined Au electrode patterns with gaps of 40~50 nm, while an AC voltage of 0.8 V at 10 MHz is applied between the gap. The samples are then cleaned with distilled water, organic solvents, and blow-dried in nitrogen before I-V measurements. The trapped colloid structure can be examined using SEM imaging after the electrical measurement. The molecular junctions made by this approach are nanoscale and addressable, but two new nanoparticle/electrode interfaces are introduced into the junction. These extra interfaces will inevitably affect the device performance and make it difficult to be analyzed. Moreover, it is unclear about the device stability because of the weak physisorption between the particle and the electrode. Similar junctions can be made by electrostatic trapping a single nanoparticle in the gap of a nanowire pattern produced by electromigration and electrodepositon methods [56,62]. There are still four interfaces in these electrode/molecule/nanoparticle/molecule/electrode junctions (Fig. 21 (middle and bottom)), although the device stability can be greatly improved because the nanoparticle is strongly bonded to the electrode through dithiol linker molecules.

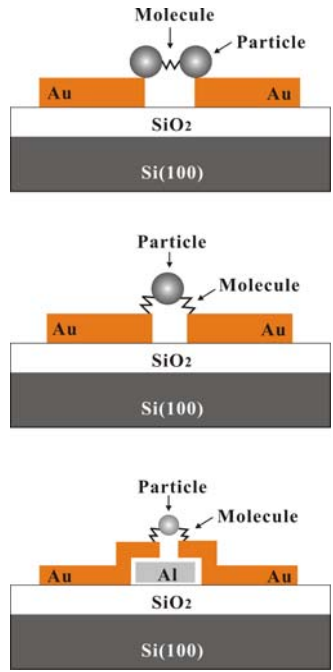

**Figure 21.** The schemes of nanoparticle bridged molecular junction with structure of (top) dimer and single particle (without or with Al gate electrode). For particle with diameter larger than 40 nm, the Au electrode pattern can be fabricated using e-beam lithography. Otherwise, the gap between the electrodes has to be further shrunk by electrodeposition or electromigration before electrostatic trapping a sub-5 nm scale nanoparticle. An aluminium film with a thin oxide layer can be used as the gate electrode in three-terminal molecular electronic device. Note that extra interfaces (either electrode/particle or particle/molecule) are introduced into these molecular junctions.

G. Liquid metal droplet based molecular junctions

A simple low-cost way to fabricate molecular junctions is the liquid metal droplet method, although the junctions are usually microscale, unstable, and not addressable. Fig. 22 shows the molecular junction with structure of liquid metal tip (Hg or GaIn), molecular layer, and base electrode on supporting substrate. A molecular junction is formed when the 'tip' and the electrode is slowly moved together using a micromanipulator. The active medium can be molecular self-assembled or Langmuir-Blodgett (LB) monolayer [33,36,101], bilayer [37,38,102], or even a thin film [39]. For bilayer junctions, the surfaces of both the tip and the electrode are covered with a SAM with the same or a different molecular structure. Multiple-layer junctions can be fabricated too. For example, an investigation of a gold nanoparticle arrays using a trilayer junction configuration: Hg-SAM/Au nanoclusters/SAM-Au has been recently reported [103]. On the other hand, the base electrode may use conductive material such as Au, Ag, Cu, ITO, doped Si, or even a Hg droplet [104,105]. Electrical characterizations are usually conducted in the presence of solutions in ambient conditions, sometimes, with the protection of Ar or $N_2$ gas. Table 4 summarizes typical results obtained on molecular junctions based on various liquid metal droplet methods, while the chemical structure for the relevant molecules are shown in Fig. 23.

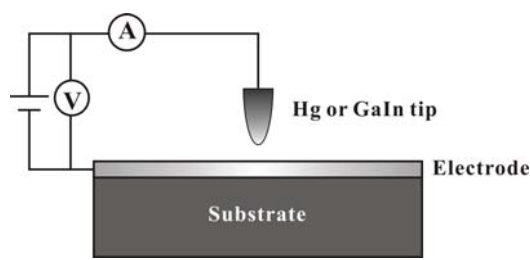

**Figure 22.** Schematic setup of liquid metal droplet based molecular junction, in which the target molecule (not shown) is assembled onto the surface of the bottom electrode and/or the top tip. Bilayer junction is formed if both electrodes are covered with monolayers. The base electrode may use Au, Ag, doped Si with or without native oxides, or even a Hg droplet. In some experiments, the junctions are measured in solution. A junction is formed when the droplet and the electrode is brought into contact with each other through slowly moving either the droplet or the substrate using micromanipulator. The bias direction could be switched back and forth for studying the charge transport mechanism.

**Table 4.** Characteristics of molecular junctions based on liquid metal droplet methods.

| Metal droplet | Device configuration | Active medium | Presence of solution | Junction area ($10^{-3}$ cm$^2$) | Bias range (V) | Ref |
|---|---|---|---|---|---|---|
| Hg | Hg-SAM/SAM-Ag | Bilayer | Yes | 4 ~ 7 | ± 1 | [37] |
|  | Hg-SAM/thin SiO$_2$/p-Si | Monolayer | No | ~ 2 | ± 0.8 | [36] |
|  | Hg-SAM/p-Si |  |  |  |  |  |
|  | Hg/SAM-SiO$_2$/p-Si |  |  |  |  |  |
|  | Hg-SAM/SAM-Au | Bilayer | Yes | ~ 2 | ± 1 | [38] |
|  | Hg-SAM/SAM-Hg | Bilayer | Yes | ~ 0.5 | -0.75 to 0.05 | [104,105] |
|  | Hg-SAM/Hg | Monolayer |  |  |  |  |
|  | Hg-SAM/SAM-Au | Bilayer | Yes | Not given | 0.05 to 0.5 | [102] |
|  | Hg-SAM/SAM protected Au nanoclusters/SAM-Au | Trilayer | No | Not given | ± 1.5 | [103] |
|  | Hg-SAM/LB/Au | Bilayer | Yes | 2.5 | ± 1 | [33] |
| GaIn | GaIn/organic film/Ag or Au, Cu, ITO | Molecular film | No | ~ 0.1 | ± 10 | [40] |
|  | GaIn/organic film/SAM-Au |  |  |  |  | [39] |

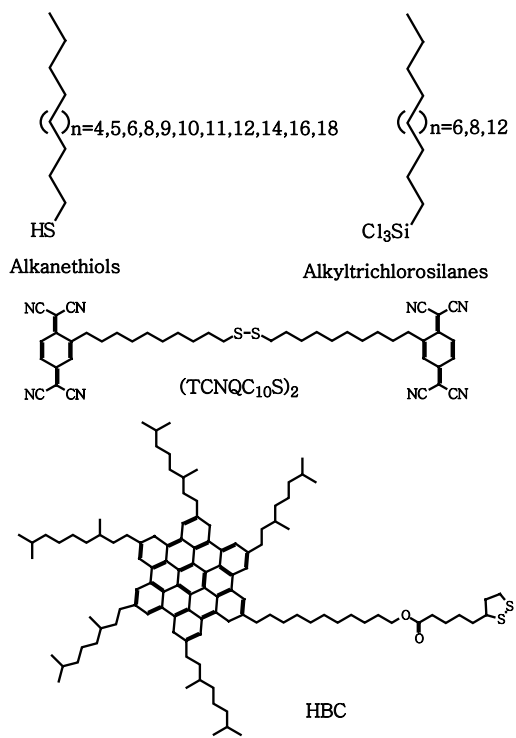

**Figure 23.** Chemical structure of molecules investigated using Hg droplet based junction approach.

Most molecular electronic element can be studied using this simple approach as long as they can form into a large-scale uniform layer on a conductive substrate. Earlier investigations have demonstrated that this method offers an effective way for fast screening molecular building blocks preferable for nanostorage. For example, the molecular first oxidation potential was found to be directly proportional to the diode turn-on voltage after a series of polyarylamines being studied using the junction setup: GaIn/molecular film/Ag [40]. It was also indicated that the arylamine chemical structure plays a key role because it affects the film morphology with larger, dendritic arylamines tending to produce smoother, more continuous films than those observed from smaller, more symmetric amines. Such dendritic arylamines may have potential for application in molecular electronic devices.

The interfaces between metal and dendritic arylamine molecular films were studied by measuring the current-voltage characteristics of single-layer organic diodes using a GaIn droplet based molecular configuration [39]. The investigation revealed that the diode turn-on voltage ($V_t$) is sensitive to the metal/arylamine interfacial electronic structure and can thus serve as a powerful probe for probing metal/molecule interface. Moreover, it was found that the junction performance strongly depends on the arylamine substituents with the cyano (–CN) groups giving a higher $V_t$ than that of the methoxy (–OCH$_3$) groups. When a monolayer of 1-decanethiol was self-assembled onto the Au anode, the diode $V_t$ is abruptly increased. Data analysis suggests that there may exist an additional energy barrier in the diodes when the arylamine with –CN groups was deposited on the SAM-Au anode. The conduction mechanism in these arylamine diodes is injection limited, which can be well described by the Richardson-Schottky (RS) thermionic emission model, because the plot of log(I) versus the square root of the effective electric field ($E_{eff}^{1/2}$) shows a linear dependence over a large range of current and voltage. A similar conduction mechanism was observed in alkyl chain monolayer junctions with structure Hg/SAM/p-Si [36], which confirms the effectiveness of the liquid metal droplet method. These results may promote future research efforts toward dendrimer based nanostorage devices and systems and this will be further discussed in the dendrimer section.

H. Charge transport mechanisms

Fig. 24 shows a schematic molecular junction with the structure of two metal electrodes bridged by a 'molecule'. The electrode/molecule interfaces may be strong chemical bond (e.g. Au–S), moderate charge-transfer interaction (e.g. Au–CN), or weak van der Waals contact. For different molecular structure and conjugation, the molecular properties may be very different. Taking these factors into account, the conductance of the molecular junction ( G ) can be expressed as [17],

$$G = T_L \cdot T_R \cdot T_M \cdot G_0$$

Where, $T_L$ and $T_R$ are the electron transmission functions at the left and right electrode/molecule contacts, respectively. While, $T_M$ is the transmission function related to the molecule itself and $G_0$ is the quantum conductance given by $G_0 = 2e^2/h$ (~77.5 μS).

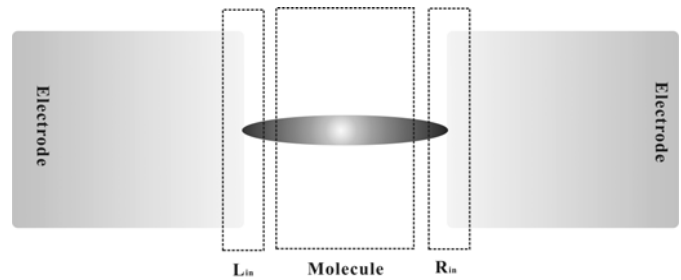

**Figure 24.** Schematic diagram of a molecular junction in which a 'molecule' is connected with two metal electrodes. As outlined by the dashed squares, serious considerations should be given not only to the molecule itself but also to the left ($L_{in}$) and the right ($R_{in}$) metal/molecule interfaces.

The charge transport through a nanoscale molecular junction is usually treated as an electron tunneling process. Based on the temperature, bias voltage, molecular properties (like conjugation degree and molecule length), it may fall into several categories: coherent nonresonant tunneling (or classic tunneling) [43,106,107,108], coherent resonant 'superexchange' tunneling [109,110], and incoherent 'diffusive' tunneling [111,112]. It has been shown that incoherent tunneling is the dominant mechanism at high temperatures, while coherent tunneling plays the main role at low temperature [113,114]. Theoretical calculations predicate that the charge transport through molecular junction may be sensitive to external optical fields [113,115,116], wire defects [112], molecular vibrational excitations [110], and electrode/molecule interfaces [112]. Some of these predictions have been proved by recent experimental results. For example, it has been demonstrated that the PDMS stamp-printing molecular junctions really respond to the stimulus of external optical signals with a different wavelength [48].

Simmons equation, as a simple model for electron tunneling through a rectangular barrier in metal/insulator/metal junctions, is widely used to analyze the experimental I-V curves [43,106]. In the low bias voltage regime, the equation can be simplified to J = $J_0$ exp (−β d) with

$$\beta = \frac{2\alpha}{\hbar}(2m\phi)^{1/2}$$

Where, J is the current density, $J_0$ a constant, β the electronic decay parameter within the molecule, d the tunneling distance, α an adjustable parameter, m electron mass, ℏ Planck's constant, and ϕ the barrier height. This simplification has been proved to be applicable in examining the effects of molecular conjugation and length on electronic transmission through molecular junctions [43,117]. When the electrode/molecule interfacial effect need to be considered, Fowler-Nordheim tunneling and Schottky-Richardson thermionic emission models are applicable in I-V data analysis for high bias voltage and high temperature regimes [20,118].

*2.2 Three-terminal molecular device*

The next milestone, after single-molecule junction toward nanostorage, may be up to realize three-terminal molecular devices [24]. For single-molecule three-terminal device, there are two configurations (Fig. 25). The first one is a molecular junction coupled with one more terminal that is not limited to gate electrode [93] and the electrolyte solution in three-terminal electrochemical cell. This extra terminal can be a tip of STM or CAFM [119], an optical or laser beam [48,113,115,120], or a nanoparticle with photoresponsive properties. The other form is more desirable and useful, whereby a three-branched molecule is connected with three electrodes through selective interaction between specific branch and designated electrode. This process can be realized by reactions of metal/sulfur and carboxyl reagent/Si with combination of functional group protecting/deprotecting techniques [68,67]. Obviously, there are still many challenges and a long way to go in the fabrication, integration, and theoretical simulation of such three-terminal devices before application in nanostorage systems.

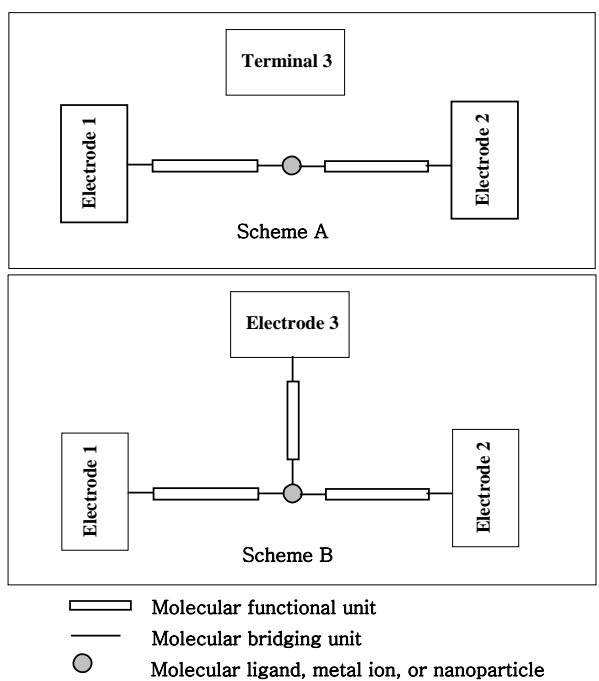

**Figure 25.** Schemes of three-terminal molecular devices. In scheme A, a single 'molecular wire' is attached to two electrodes, in which the third terminal can be a tip of STM or CAFM, a coupled nanoparticle, or a laser beam. In scheme B, a branched 'molecule' is connected with three electrodes through selective connection of each specific branch with designated electrode.

*2.3 Dendrimer based memory devices*

Dendrimers have become increasingly important in recent years because of their potential applications in information storage [121,122,123], electro/optical switches [124,125,126,127,128,129], sensors [130,131,132], solar cells, and light emitting diodes [133,134,135,136]. Dendrimers may be classified into two main types (Fig. 26). The big advantage of dendrimers is related to their core/shell structure, nanoscale dimensions, and the presence of many functional sites and the controllability of these sites [134]. The electrical and/optical properties of dendrimers can be thus tuned by changing (1) the core unit, (2) the chemical nature of the shell ligands and/or terminal groups, and (3) by complexation with different nanoparticles and/or metal ions [137,138,139,140]. Such capability makes dendrimers one of the most flexible candidates for building blocks in bottom-up nanostorage system.

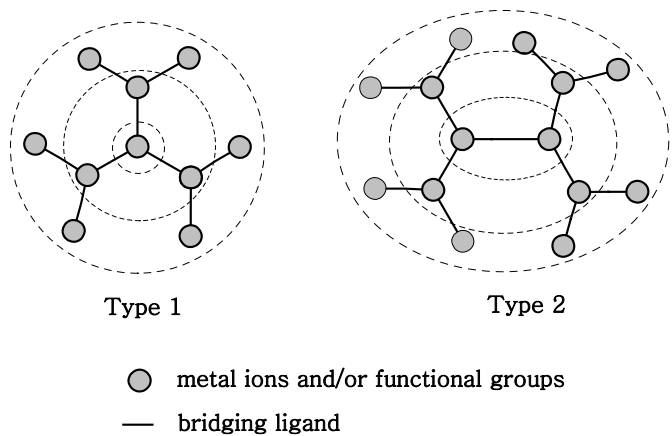

**Figure 26.** Schematic structures usually used in dendrimers. The dashed circles outline the core/shell(s) structure.

Ultraviolet light can induce reversible *cis/trans* isomerization reactions in compounds with an azobenzene moiety (Fig. 27). Various dendrimers containing such light switchable units have been synthesized. As shown in Fig. 28, one choice is to use the azobenzene unit as the central linker (**1**) [124,133], the other approach is to place it in the periphery (**2**, **3**) [126,127,129]. Redox-switchable dendrimers incorporating both electron donor and acceptor groups such as TTF and cyanobiphenyl have been studied [125,128]. Other kinds of dendrimers complexed with metal ions (such as $Sn^{2+}$, $Ru^{2+}$, $Fe^{2+}$, $Rh^{3+}$, $Ir^{3+}$ and $Os^{2+}$) [121,130,131,134,137,138] and/or nanoparticles [139,140] were reported too. Most of the dendrimers were investigated in the form of either molecular solutions [124,129] or monolayers at the air/water interface [126,127]. However, much of the interest in these dendrimers is to know if similar performance can be obtained in solid-state single-dendrimer devices. A critical step is to study the controlled assembly onto surfaces of different conducting substrates (such as Au, Ag, Pd, and Pt).

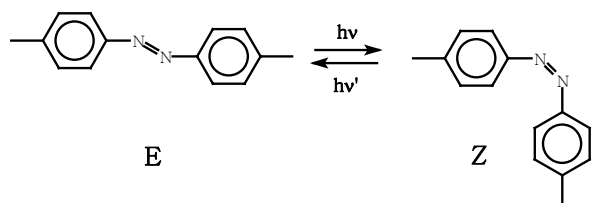

**Figure 27.** Ultraviolet light induced reversible *cis/trans* isomerization of azobenzene moiety.

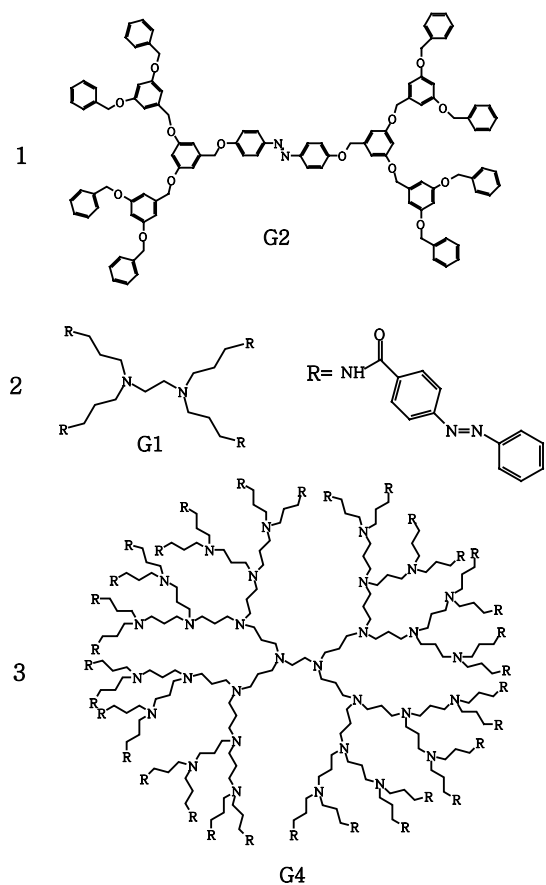

**Figure 28.** Photoresponsive dendrimers with zaobenzene as central linker (**1**) and periphery groups (**2**,**3**), respectively [124,129]. Molecule **1** is the 2nd generation dendrimer (G2), while **2** and **3** are the 1st (G1) and 4th (G4) generations, respectively.

Drop-casting, spin-coating, or vapor deposition techniques can be used to prepare dendrimer thin films on various substrates. It has been shown that most dendrimer films are in the amorphous state and composed of small island-shaped domains due to their large, symmetric, globular, and dense molecular structure [39,40,141]. On the other hand, PDMS stamp-printing, solution adsorption, and LB methods can be employed to fabricate multilayer or even monolayer dendrimers. However, there normally exists a large amount of patches/defects in the layer due to the effects of physical contact, solvent, and/or layer transfer [142]. This may inevitably result in high short-circuit rates in the final devices. The traditional SAM approach, as used in small thiolated molecules, might be a good choice. A thiol-derivatized-alkyl linker can be added to the periphery of dendrimers. Moreover, to largely reduce the SAM defects, an alkanethiol matrix is necessary for inserting individual dendrimers. The methods for fabrication of solid-state dendrimer layers are summarized in Table 5.

**Table 5.** Approaches to fabricate dendrimer layer.

| Fabrication approach | Layer | For nanostorage | Refs. |
|---|---|---|---|
| Drop-casting | Film to multilayer | No | [140,143] |
| Spin-coating | Film to multiplayer | No | [130,136] |
| Vapor deposition | Film to multiplayer | No | [40,144] |
| Stamp-printing | Multi to monolayer | Possible | [145] |

| Solution adsorption | Multi to monolayer | Possible | [146] |
| --- | --- | --- | --- |
| LB monolayer | Monolayer | Possible | [126] |
| SAMs | Monolayer | Very possible | [122] |

Redox-active dendritic molecules may be promising candidates for molecular-scale charge storage. A series of zinc porphyrins has been investigated through electrochemical measurements, in which each porphyrin bears a S-acetylthiol-derivatized linker and can thus form SAM onto Au electrodes [122,123] (Fig. 29). The gold microelectrode with porphyrin SAM was measured in a two-terminal electrochemical cell with dried and distilled $CH_2Cl_2$ containing 0.1 M electrolyte of $Bu_4NPF_6$. Information was stored via removing electrons from the porphyrin units by applying a threshold bias voltage between the electrodes above the molecular oxidation potential. It was shown that multiple oxidation states (i.e. neutral, monocation, and dication) of the porphyrins can be reversibly obtained. The charge retention time is hundreds of seconds. Similar redox-active information storage was also observed in electrochemical experiments of dendritic molecules such as 4AA/PD and CN-4AAPD [39,40] (Fig. 29). The molecular layers were vapor deposited onto Au electrodes. Oxidation induced color change was observed in the films, which were stable for more than 3 months in ambient conditions. However, a question is what roles the electrolyte and solvent play in these works.

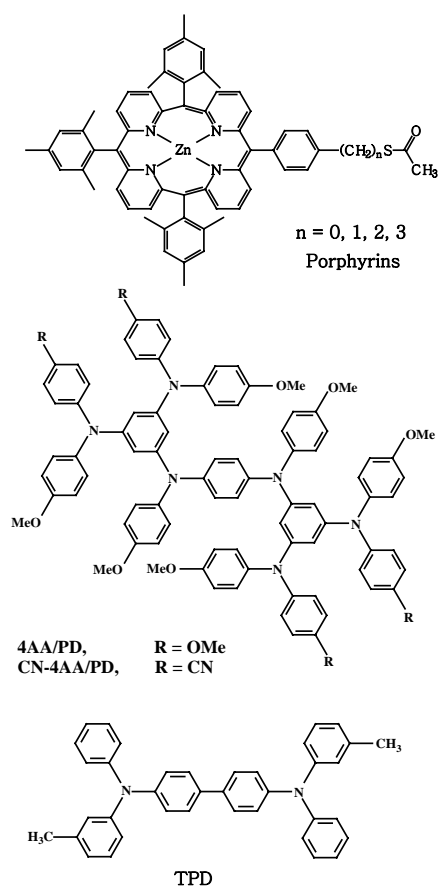

**Figure 29.** Chemical structures of redox-active dendritic porphyrins, 4AA/PD and CN-4AA/PD. The molecular structure of TPD is also shown, which is well known as a good hole transport material for organic light emitting diodes.

To address this issue, the electrical performance of 4AA/PD thin films was studied in device configuration without organic solvent or electrolyte [147]. The shadow mask assisted crossbar approach was used to construct junctions of Ag/4AAPD/Ag, Ag/TPD/Ag, Ag/4AAPD/TPD/Ag, and Ag/TPD/4AAPD/TPD/Ag. The films were vapor deposited in high vacuum condition and were immediately measured after removing from the chamber. It was shown that no such redox charge storage behavior exists in solid-state devices of Ag/4AAPD/Ag, Ag/TPD/Ag or Ag/4AAPD/TPD/Ag in spite of the variation of layer thickness and metal material. However, irreversible electrical switching was observed from devices with the structure: Ag/TPD/4AAPD/TPD/Ag. Control experiment suggested that it is due to charge storage in the 4AAPD molecular layer sandwiched by the TPD thin films. Since the oxidation potential of TPD is higher than that of the 4AAPD, an energy well may thus form in the sandwiched 4AA/PD film and where charges can be trapped [40]. This result may be applicable to design dendrimers with optimized core/shell structure for molecular-scale charge storage.

## 3  Bioelectronics

The advance of nanostorage has a close relationship to the field of bioelectronics. Because one main object of bioelectronics is to find effective ways to develop novel nanoscale devices, building blocks, principles, and technologies for ultrahigh density, low-cost and intelligent information storage systems. The basic concept of bioelectronics is to address (electrically, optically, and/or magnetically) a single biomaterial immobilized on substrate. The biomaterials may include proteins, DNA, and enzymes [148,149], which can be treated as either an organic molecule or nanomaterial. The assembly of these materials can be realized using covalent bond (e.g. Au-thiol), affinity interaction (e.g. charge transfer), and hydrophilic/hydrophobic interaction. For example, Albert suggested in 1968 that charge transfer should be one of the most common and fundamental biological reactions and it may play a key role in biological regulation, defence, and cancer [22].

Great advances have been made in biomaterial synthesis, layer assembly, fabrication of electrical contacts, and tailoring the biomaterial/electrode interfaces. Willner and co-workers reviewed the integration of layered biomaterials and conductive substrates for application in biosensors, bioelectronic arrays, logic gates, and optical memories [148,150]. They also reviewed the development of nanoparticle-biomolecule hybrid systems from the perspective of synthesis, properties, and applications [151]. DNA, as a chemically based molecular system, may be a key player in the bottom-up approach of nanotechnology owing to its features of about 2 nm-diameter, 4 nm helical repeat, and 50 nm persistence length [152]. It was suggested that DNA can be used as conductive nanowires, branched nanoarchitectures, computers, nanomachines, and sensors [153,154].

These studies were mostly focused on the synthesis, organization/immobilization, modification, and simple characterization of the biomaterials [148]. Although, it was realized to controllably deposit molecules of interest to desired locations on substrate using AFM based dip-pen nanolithography [90] and/or stamp printing methods [91]. Electrical properties of single biomaterial were measured using CAFM or STM in combination with nanoparticles [151]. As discussed in former sections, in order to reproducibly address a single biomolecule and to further integrate it into a functional system, reliable methods still need to be established whereby sub-5 nm scale metal-molecule-metal junctions can be efficiently made. Moreover, the biomaterial/electrode interfaces need to be fully understood and carefully engineered. From the viewpoint of device fabrication and characterization, the molecular junction approaches are applicable to bioelectronics [21]. More efforts are needed to address the development of hybrid nanoscale device and information system with combinations of molecules, biomaterials, and nanostructures. It might be possible to realize semi-intelligent information storage systems through mimicking the human brain functions and structures [155,156]. The first step is to build suitable neuron models by analyzing the biological neuron systems and their functions [157].

With such models and understanding, novel information devices and technologies may be available in the 'near' future.

## 4 Nanoelectronics

Nanoparticles are also known as quantum dots, nanodots, nanocrystals, or nanograins. Basically, they can be treated as man-made 'molecule'. A variety of nanoparticles have been fabricated by methods of chemical reaction [158], lithography [159,160], chemical vapor deposition (CVD) [161,162], deposition plus annealing [163,164], molecular-beam epitaxy (MBE) [165], and anodization plus rapid thermal oxidation [166]. Nanoparticles, with sub-5 nm dimension, are believed to be one of the most promising materials for ultrahigh density memory devices [167]. For example, nanoparticles covered with insulating layers have been investigated for applications in nonvolatile flash memory devices [168] and electrostatic data storage [9,169,170]. The principle of the memory device is to inject a number of charges into a single or bundle of nanoparticle embedded in insulator (i.e., WRITE), store the injected charges (STORAGE), and later to sense the stored charges (i.e., READ). The data erasing is realized by removing the stored charges using either reversed electrical field and/or optical illumination. This charge injection-storage-erase process requires the nanoparticle to possess discrete energy levels where the charges can stay, which was confirmed by the electrical measurements of Au nanoparticles at the temperature of 12 K [171,172,173,174]. Semiconductor nanoparticles like Si, Ge, CdSe, InAs, and Ga(In)As should be the most promising candidates to be examined for information storage [158,160,175,176,177].

Nanocrystalline silicon (nc-Si) has been studied for application in single-electron memory device [160,178,179]. However, it was argued that nanodots of Ge [180,181], refractory metal (like TiN and W), and SiGe [162,182] should be more suitable for nanomemory application owing to their smaller band gap than that of silicon. Other kinds of information memory and processing systems were also investigated that include quantum-dot based cellular automata [183,184], electron-spin memory [185,186], and single-electron parametron [187].

Although there are some breakthroughs in nanoparticle based memory device, major challenges still remain. The following issues need be addressed:

1) Most of the nanoparticle based memory devices are currently made by photolithography or e-beam lithography. As a result, the device dimension is still too large to be used for nanostorage. Low-cost fabrication approaches are needed to make large array of sub-5 nm single nanoparticle device.
2) The tunneling barrier is not well defined. Effects of the material, thickness and structure of the embedding insulating layer (or organic ligands) have to be examined in detail.
3) The device operating temperature is currently too low, which need to be improved as close as possible to room temperature.
4) More effort on controlling the size, shape, uniformity, and assembly of nanoparticles.
5) Characterizing the electrical properties (e.g. I-V curves) of single-nanoparticle device in the presence of optical and/or magnetic fields.
6) To understand (or even to control) how the nanocrystal structure affect the device performance.

Nanotubes and nanowires are also being extensively investigated for application in various nanoscale devices [2,13,188]. Besides nanowires/tubes can be used to construct addressable molecular junction arrays [86,89], these nanostructures themselves can be used as information storage media [11,16]. For example, electrical switching has been reported for carbon nanotube Y-junctions [189]. Single-nanowire superlattice with multi-heterojunctions is also synthesized and characterized [14,15]. More efforts are necessary to develop some low-cost methods to controllably fabricate nanostorage device array [190].

32    *J. C. Li*## 5  Summary

The development of nanoscale data storage has been reviewed from the viewpoints of molecular device fabrication and the building blocks used. It is shown that nanostorage device should be sub-5 nm scale, addressable, durable, low-cost, highly integrated, and compatible with bioelectronics and/or nanoelectronics. It seems quite possible to achieve applicable nanostorage devices using crossed nanowire/nanotube approaches in combination with the techniques of electrodeposition, self-assembly, nanoparticle assembly and bioelectronics. Currently, it is still too early to conclude which approaches and building units will be ultimately used in the 'coming' commercial nanostorage devices. As one of the most critical steps, more effort should be made to figure out some simple but applicable methods for fabrication of nanoscale devices. In the mean time, the current approaches can be used as test-beds to investigate various building blocks from aspects of molecular synthesis, self-assembly, and effects of molecule/electrode interfaces, chemical structure, functional groups and/or local environment on the device performance. Moreover, we may ask what is next after the generation of nanostorage? How will it affect our daily life? One answer might be semi-intellectual memory systems with some characteristics of the human brain, which may be achieved someday using the novel technologies, equipments, materials, and theories developed on the basis of nanostorage, molecular electronics, nanoelectronics, and bioelectronics.

**Acknowledgements**

JCL thanks Prof. Paul Mulvaney for his suggestions in preparing this manuscript and acknowledges the Australia ARC for support under DP Grant 0558608.

**Dr. Jianchang Li** was born in 1971 in Hebei province, China. He received his BS and MS degrees in Vacuum Science and Technology from Northeastern University, Shenyang, in 1994 and 1997, respectively. He completed his PhD studies in Physical Electronics in 2000 at Peking University, Beijing. After one year postdoctoral stay at City University of Hong Kong, he moved to The University of Alabama, Tuscaloosa, in 2001 and worked as a postdoctoral research fellow at The Center for Materials for Information Technology for three years. In 2004 he joined The University of Chicago as a research associate. He is currently working at The University of Melbourne, Australia, as a research fellow beginning at 2006. His research interests include surfaces and interfaces, thin films and monolayers, vacuum technologies, molecular electronics, and nanoelectronics.